# Publishing Microdata with a Robust Privacy Guarantee[*]

Jianneng Cao
School of Computing
National University of Singapore
caojn@purdue.edu

Panagiotis Karras
Management Science and Information Systems
Rutgers University
karras@business.rutgers.edu

## ABSTRACT

Today, the publication of microdata poses a privacy threat. Vast research has striven to define the privacy condition that microdata should satisfy before it is released, and devise algorithms to anonymize the data so as to achieve this condition. Yet, no method proposed to date explicitly bounds the percentage of information an adversary *gains* after seeing the published data for *each* sensitive value therein. This paper introduces $\beta$-likeness, an appropriately robust privacy model for microdata anonymization, along with two anonymization schemes designed therefor, the one based on generalization, and the other based on perturbation. Our model postulates that an adversary's confidence on the likelihood of a certain sensitive-attribute (SA) value should not increase, in relative difference terms, by more than a predefined threshold. Our techniques aim to satisfy a given $\beta$ threshold with little information loss. We experimentally demonstrate that (i) our model provides an effective privacy guarantee in a way that predecessor models cannot, (ii) our generalization scheme is more effective and efficient in its task than methods adapting algorithms for the k-anonymity model, and (iii) our perturbation method outperforms a baseline approach. Moreover, we discuss in detail the resistance of our model and methods to attacks proposed in previous research.

## 1. INTRODUCTION

Organizations, such as government agencies or hospitals, regularly release microdata (e.g., census data or medical records) to serve benign purposes. However, such data can inadvertently reveal sensitive personal information to malicious adversaries. Experience has shown that merely concealing explicit identifying attributes, such as *name* or *phone number*, does not suffice to protect personal privacy. An attacker may still uncover hidden identities and/or sensitive information, by joining the released microdata attributes with other publicly available data. The set of attributes instrumental to that purpose, such as *gender*, *zipcode*, and *age*, are called *quasi-identifiers* (*QI*s). The anonymization problem calls for bringing the data to a form that forestalls such linking attacks while preserving as much of the original information as possible.

The question of the form the data should be brought to is a subject of inquiry in itself. Past research has tried to formulate a *privacy guarantee* an anonymized data set should satisfy, using syntactic and perturbation-based methods.

Syntactic anonymization methods typically postulate that microdata be partitioned into a set of *equivalence classes* (ECs), such that all tuples within an EC be indistinguishable from (or mutually interchangeable with [33]) each other as far as their *QI*s are concerned. The models differ in the condition that an eligible EC should satisfy. By $k$-anonymity, each EC should consist of at least $k$ tuples [29]. In effect, $k$-anonymity protects against *identity disclosure*, as it hides each released tuple in a crowd of at least $k-1$ others, but does not attend to the values of a non-*QI sensitive attribute* ($\mathcal{SA}$); hence, the privacy regarding such values may be compromised. To address this limitation, $\ell$-diversity requires that each EC contain at least $\ell$ different "well represented" $\mathcal{SA}$ values (in a mathematical sense) [22]. Even so, $\ell$-diversity fails to protect against attacks arising from an adversary's unavoidable knowledge of each $\mathcal{SA}$ value's frequency in a released table. As a rectification to this problem, $t$-closeness proposes a condition that bounds the cumulative difference between the frequency distribution of $\mathcal{SA}$ values in an EC and their overall distribution [20]. Yet, as we will discuss, such a bound fails to provide a meaningful privacy guarantee that lays grounds for effective and human-understandable policy [25].

Perturbation-based methods add noise to the data so as to achieve a privacy property. The models in [10, 30, 5] impose a bound on an adversary's *posterior* confidence about a data property in relation to the prior one; however, they measure confidence gain in *absolute*, not in *relative* terms. Other noise-adding methods enforce differential privacy [9], which guarantees that the effect of any particular individual's data on a query result is dominated by the noise; in other words, the result is broadly the same, regardless of whether a certain individual has contributed her true information. Yet, as [6] shows, an individual's $\mathcal{SA}$ value can be inferred from differentially private data with non-trivial accuracy, while the added noise can dominate small values in the results of aggregate queries [32].

In this paper, we propose $\beta$-likeness: a robust and intuitive model for microdata anonymization, postulating that an adversary's confidence in a tuple's $\mathcal{SA}$ value should not increase in *relative* terms by more than a threshold after seeing the published data. We accompany this model with two anonymization schemes tailored for its particular requirements: one based on generalization, and one on perturbation; the latter can better handle remote outliers. We experimentally demonstrate that our schemes: (i) provide effective privacy guarantees in a way that state-of-the-art $t$-closeness schemes cannot; and (ii) are more efficient than competing approaches.

---

[*]Work supported by a Rutgers Business School RRC grant.

## 2. RELATED WORK AND ARGUMENT

The first model suggested for anonymizing microdata while preserving their integrity was $k$-anonymity [29]; it suggests grouping tuples in ECs of at least $k$ tuples each, with indistinguishable $QI$ values. As the problem of *optimal* (i.e., minimum-information-loss) $k$-anonymization is NP-hard [23] in non-trivial cases, past research has proposed several heuristics. Such schemes transform the data by *generalization* and/or *suppression*. Generalization replaces, or *recodes*, all values of a $QI$ attribute in an EC by a *range* containing them. For example, $QI$ gender with values *male* and *female* can be generalized to *person*, and $QI$ age with values 20, 25 and 32 can be generalized to [20, 32]. Suppression is an extreme case of generalization that deletes some $QI$ values or even tuples.

Still, the $k$-anonymity model suffers from a critical limitation. While its objective is to conceal sensitive information, it pays no attention to non-$QI$ *sensitive* attributes ($\mathcal{SA}$s). A $k$-anonymized table may contain ECs with so skewed a distribution of $\mathcal{SA}$ values, that an adversary can still infer the $\mathcal{SA}$ value of a record with high confidence. To address this limitation, [22] proposed $\ell$-diversity, which postulates that each EC contain at least $\ell$ "well represented" $\mathcal{SA}$ values, where "well represented" can be defined in diverse ways.

Still, $\ell$-diversity fails to guarantee privacy when the *distribution* of $\mathcal{SA}$ values differs substantially among ECs and from their *overall* distribution; thus, it is vulnerable to a *skewness* attack [20]. For instance, assume a 10-diverse form $\mathcal{T}'$ of a medical record table $\mathcal{T}$, in which $0.1\%$ persons are infected with HIV, and an EC $\mathcal{G} \in \mathcal{T}'$ containing 10 distinct $\mathcal{SA}$ values, with one occurrence of HIV. The probability of HIV is $10\%$ for a tuple in $\mathcal{G}$, but only $0.1\%$ for a tuple in $\mathcal{T}$. This 100-fold increase of probability is a significant, hence undesirable, information leak. Furthermore, a *similarity* attack [20] is likely when the $\mathcal{SA}$ values in an EC are semantically similar. For example, a 3-diverse table can be generated from Table 1 by putting the first 3 tuples in EC $\mathcal{G}_1$, and the rest EC $\mathcal{G}_2$. Regardless of their diversity, all tuples in $\mathcal{G}_1$ indicate a *nervous problem*.

| ID | Name | Weight | Age | Disease |
|----|------|--------|-----|---------|
| 01 | Mike | 70 | 40 | headache |
| 02 | John | 60 | 60 | epilepsy |
| 03 | Bob | 50 | 50 | brain tumors |
| 04 | Alice | 70 | 50 | heart murmur |
| 05 | Beth | 80 | 50 | anemia |
| 06 | Carol | 60 | 70 | angina |

**Table 1: Patient records**

To forestall these attacks, Li et al. proposed $t$-closeness, which requires that a *cumulative* difference of the $\mathcal{SA}$ values' distribution within any EC from the one in the overall table does not exceed a given threshold $t$ [20]. The $t$ threshold is meant to constrain the information an adversary gains after seeing a single EC, with respect to that provided by the full released table. Just like $\ell$-diversity is open to many ways of measuring the number of "well-represented" values in an EC [22], the $t$-closeness model is open to diverse ways of measuring the cumulative difference between the overall $\mathcal{SA}$ distribution, $\mathcal{P}$, and that in an EC, $\mathcal{Q}$. One option is the Earth Mover's Distance (EMD) [28]. Another proposal [20] first transforms $\mathcal{P}$ ($\mathcal{Q}$) to $\widehat{\mathcal{P}}$ ($\widehat{\mathcal{Q}}$) by kernel smoothing, and then calculates the Jensen-Shannon divergence between $\widehat{\mathcal{P}}$ and $\widehat{\mathcal{Q}}$ as the approximate distance between $\mathcal{P}$ and $\mathcal{Q}$. Last, the Kullback-Leibler divergence is used in [27]. Yet these functions all interpret the $t$ threshold as a bound on the *cumulative* difference between two frequency distributions. Indeed, this interpretation emanates out of the $t$-closeness model itself [20]. Still, a privacy model should provide grounds for effective and human-understandable policy [25]. Models that bound a *cumulative* function of frequency differences between distributions fails to provide a comprehensible relationship between the $t$ threshold and the privacy it affords. In particular, such models do not pay due attention to *less frequent* $\mathcal{SA}$ values, which are more vulnerable to privacy exposure; and do not distinguish between positive and negative variation in an $\mathcal{SA}$ value's frequency.

We first elaborate on EMD. Assume a data set $\mathcal{DB}$ with $\mathcal{SA}$ values HIV and Flu. If the overall $\mathcal{SA}$ distribution between them is $\mathcal{P} = (0.4, 0.6)$, and their distribution in an EC is $\mathcal{Q} = (0.5, 0.5)$, then $EMD(\mathcal{P}, \mathcal{Q}) = 0.1$. Still, if their overall distribution is $\mathcal{P}' = (0.01, 0.99)$ and their distribution in an EC is $\mathcal{Q}' = (0.11, 0.89)$, then $EMD(\mathcal{P}', \mathcal{Q}') = 0.1$ again. Both cases satisfy 0.1-closeness. However, the information gain in the latter case is much larger than that in the former: the probability of HIV rises by $25\%$ from 0.4 to 0.5, but by $1000\%$ from 0.01 to 0.11. In effect, the two cases do *not* afford the same privacy. This example appears in [20], where it is noted that EMD does not provide a clear privacy guarantee. In fact, not only EMD, but any function that *aggregates absolute differences* faces a similar problem, since such functions do not provide *maximum relative difference* guarantees [14, 13] about individual $\mathcal{SA}$ values. In our example, a small relative difference of Flu-frequency evens up a large relative difference of HIV-frequency.

K-L divergence [27] and J-S divergence [20, 21] also fail to pay *equal* attention to *all* $\mathcal{SA}$ values and their relative differences. In our running example, assume a dataset where the overall distribution of HIV and Flu is $\widetilde{\mathcal{P}} = (0.01, 0.99)$, and their distribution in an EC is $\widetilde{\mathcal{Q}} = (0.03, 0.97)$. Then the K-L (J-S) divergence between $\mathcal{P}$ and $\mathcal{Q}$, is 0.0290 (0.0073), while that between $\widetilde{\mathcal{P}}$ and $\widetilde{\mathcal{Q}}$ is 0.0133 (0.0038). Both these alternatives estimate the privacy afforded by $\widetilde{\mathcal{Q}}$ with respect to $\widetilde{\mathcal{P}}$ as *higher* than that afforded by $\mathcal{Q}$ with respect to $\mathcal{P}$. However, the confidence for HIV increases only by $25\%$ in the latter case, while it rises by $200\%$ in the former.

Besides, the anonymization schemes in [20] are mere extensions of $k$-anonymization techniques [17, 18]. They do not cater to the special needs of $t$-closeness, hence yield low information quality. Recently, [4] proposed an anonymization algorithm specialized for $t$-closeness, yet did not discuss the limitations of the model itself. Last, the anonymization scheme in [27] uses perturbation and adds noise to the data, damaging their truthfulness.

The privacy model of [10] imposes a bound $\rho_2$ to the *posterior* probability (i.e., after release) of certain properties in the data, given a bound $\rho_1$ on the *prior* probability (i.e., before release). This model is modified in [30], where the posterior confidence should not exceed the prior one by more than $\Delta$. These models measure the *absolute* confidence gain (i.e., information leak), hence do not sufficiently protect the privacy of infrequent values. For example they treat a probability increase from $60\%$ to $80\%$ as tantamount to an increase from $1\%$ to $21\%$ in *absolute* terms, while the latter is an increase by $2000\%$ and the former by $33\%$ in *relative* terms.

Alternative approaches enforce differential privacy [9]. By this model, the data owner adds noise to a query result so as to guarantee that this noisy result would change very little with the variation of a particular individual's data. However, [16] illustrates that differential privacy does not adequately limit inference about an individual's participation in the data generating process. Furthermore, and more importantly for the focus of our work, [6] has recently shown that, even though the effect of any single individual is dominated by the added noise, the noise itself is *in turn dominated by the signal emerging from the whole population*. Consequently, one can effectively build a Naïve Bayes classifier inferring individuals' $\mathcal{SA}$ values with non-trivial accuracy [6].

A recently proposed distribution-oriented privacy model is $\delta$-disclosure-privacy [3]; it requires that for any $\mathcal{SA}$ value $v_i$ with frequency $p_i$ in the original table, its frequency in any EC, $q_i$, should be such that $\left|\log\left(\frac{q_i}{p_i}\right)\right| < \delta$. Yet this model fails in two respects:

(1) since $\log(q_i)$ is defined only for $q_i > 0$, $\delta$-disclosure-privacy strictly requires that each $\mathcal{SA}$ value in the original table occurs in every EC; (2) given a sufficiently large value of $p_i$ and a modest value of $\delta$, $\delta$-disclosure-privacy does not effectively upper-bound $q_i$, hence allows for absolute certainty of one's SA value, which is exactly the kind of leak it is meant to prevent. These properties render $\delta$-disclosure-privacy unnecessarily rigid, in one way, and yet exceedingly lax, in another way. Besides, [3] does *not* propose an anonymization algorithm tailored for $\delta$-disclosure-privacy; it only points out that the Mondrian $k$-anonymization algorithm [18], adapted for $\delta$-disclosure-privacy (as well as for $\ell$-diversity and $t$-closeness), yields high information loss. This *negative* result is not surprising; after all, Mondrian simply partitions the data to disjoint ECs, hence is ill-suited for models looking into the sensitive values in an EC, as observed in [12] In its conclusions, [3] observes that better anonymization algorithms are needed for those models, but does not provide such algorithms; it focuses on a *negative* result without attempting to ameliorate it. In this paper, we provide a meaningful distribution-oriented privacy model that avoids the drawbacks of $\delta$-disclosure-privacy and $t$-closeness, *as a well as* an anonymization algorithm specifically designed therefor. Thus, our work goes beyond [3] in all these respects.

## 3. THE PRIVACY MODEL

This section introduces our privacy model. Our model assumes that the $\mathcal{SA}$ distribution in $\mathcal{DB}$ is public knowledge, and constrains the $\mathcal{SA}$-related information gained by the table's publication. Table 2 gathers together the notations we use.

| | |
|---|---|
| $\mathcal{DB}$ | Original microdata table |
| $\mathcal{SA}$ | Sensitive attribute in $\mathcal{DB}$ |
| $\mathcal{V} = \{v_1, v_2, \ldots, v_m\}$ | The domain of $\mathcal{SA}$ |
| $N_i$ | Number of tuples with $v_i$ in $\mathcal{DB}$ |
| $p_i = N_i/|\mathcal{DB}|$ | Frequency of $v_i$ in $\mathcal{DB}$ |
| $\mathcal{P} = (p_1, p_2, \ldots, p_m)$ | Overall $\mathcal{SA}$ distribution in $\mathcal{DB}$ |
| $\mathcal{G}$ | Equivalence class |
| $\mathcal{Q} = (q_1, q_2, \ldots, q_m)$ | $\mathcal{SA}$ distribution in $\mathcal{G}$ |

**Table 2: Notations**

DEFINITION 1 (INFORMATION GAIN). *Assume that $\mathcal{DB}$ is a table with a sensitive attribute $\mathcal{SA}$. Let $\mathcal{V} = \{v_1, v_2, \ldots, v_m\}$ be the $\mathcal{SA}$ domain, and $\mathcal{P} = (p_1, p_2, \ldots, p_m)$ be the overall $\mathcal{SA}$ distribution in $\mathcal{DB}$. Suppose that $\mathcal{Q} = (q_1, q_2, \ldots, q_m)$ is the $\mathcal{SA}$ distribution in an equivalence class $\mathcal{G}$, formed by tuples from $\mathcal{DB}$. The* information gain *on any $\mathcal{SA}$ value $v_i \in \mathcal{V}$ is $\mathsf{D}(p_i, q_i)$, where $\mathsf{D}$ is a distance function between $p_i$ and $q_i$.*

We say that the information gain on $v_i$ is *positive*, when $p_i < q_i$, and *negative*, when $p_i \geq q_i$. Negative information gain lowers the correlation between a personal record and $v_i$ in EC $\mathcal{G}$ below that in the whole table. In most cases, such gain enhances privacy. However, there may exist $\mathcal{SA}$ values such as *heterosexual*, for which a reduced likelihood may inadvertently violate privacy. Nevertheless, we assume that the $\mathcal{SA}$ domain always includes the negation of such values. Thus, negative information gain on *heterosexual* always appears as positive gain for *homosexual*. Therefore, we can directly control the positive gain on the value (such as *homosexual*) that poses the privacy threat. For a more general case such as *marital status*, the negative gain on $\mathcal{SA}$ value *married* can imply that an individual is more likely to be *divorced* or *widowed*. However, we assume that the $\mathcal{SA}$ domain contains all the values of interest. Hence, the relative negative gain of *married* can be transformed to the positive gains of *divorced*, and *widowed*. Based on the above reasonable assumption, we are concerned with *positive* information gain; negative gain *can* be treated symmetrically if circumstances demand it (see Section 7). We define basic $\beta$-likeness as follows.

DEFINITION 2 (BASIC $\beta$-LIKENESS). *Given table $\mathcal{DB}$ with sensitive attribute $\mathcal{SA}$, let $\mathcal{V} = \{v_1, \ldots, v_m\}$ be the $\mathcal{SA}$ domain, and $\mathcal{P} = (p_1, \ldots, p_m)$ the overall $\mathcal{SA}$ distribution in $\mathcal{DB}$. An EC $\mathcal{G}$ with $\mathcal{SA}$ distribution $\mathcal{Q} = (q_1, \ldots, q_m)$ is said to satisfy basic $\beta$-likeness, if and only if $\max\{\mathsf{D}(p_i, q_i) | p_i \in \mathcal{P}, p_i < q_i\} \leq \beta$, where $\beta > 0$ is a threshold.*

For a table $\mathcal{DB}'$ anonymized from table $\mathcal{DB}$ to obey $\beta$-likeness, all equivalence classes $\mathcal{G} \subset \mathcal{DB}'$ have to conform to $\beta$-likeness. Contrary to previous models [20, 3, 21, 27], basic $\beta$-likeness clearly quantifies the relationship between the $\beta$ threshold and positive information gain. Thanks to the maximum-distance threshold it imposes, it inherently safeguards against *skewness attacks* and *semantic attacks* [20]. Last, as it clearly distinguishes between positive and negative information gain, and accepts $\mathcal{SA}$ values absent from an EC, it allows for more flexibility in anonymization, hence higher information quality, than the closest related model, $\delta$-disclosure-privacy [3]. Apart from specifying a *maximum*, instead of a cumulative, distance threshold, we should also define the distance function $\mathsf{D}$ in an appropriate manner. As we have argued, a measure of *absolute* difference does not serve our purposes, since it fails to protect less frequent $\mathcal{SA}$ values. We opt for *relative* difference instead, and define the distance function as $\mathsf{D}(p_i, q_i) = \frac{q_i - p_i}{p_i}$. This function obeys the *monotonicity property*.

LEMMA 1 (MONOTONICITY PROPERTY). *Assume that $\mathcal{SA}$ value $v_i \in \mathcal{V}$ has frequency $p_i$ in the overall table $\mathcal{DB}$, $q_i^1$ ($q_i^2$) in EC $\mathcal{G}_1$ ($\mathcal{G}_2$), generated from tuples in $\mathcal{DB}$, and $q_i^3$ in $\mathcal{G}_1 \cup \mathcal{G}_2$. Then $\mathsf{D}(p_i, q_i^3) \leq \max\{\mathsf{D}(p_i, q_i^1), \mathsf{D}(p_i, q_i^2)\}$.*

PROOF. Assume there are $n_1$ ($n_2$) tuples with $v_i$ in $\mathcal{G}_1$ ($\mathcal{G}_2$). Then $q_i^1 = \frac{n_1}{|\mathcal{G}_1|}$, $q_i^2 = \frac{n_2}{|\mathcal{G}_2|}$, $q_i^3 = \frac{n_1+n_2}{|\mathcal{G}_1|+|\mathcal{G}_2|} = \frac{q_i^1|\mathcal{G}_1|+q_i^2|\mathcal{G}_2|}{|\mathcal{G}_1|+|\mathcal{G}_2|} \leq \max\{q_i^1, q_i^2\}$. Thus, $\mathsf{D}(p_i, q_i^3) \leq \max\{\mathsf{D}(p_i, q_i^1), \mathsf{D}(p_i, q_i^2)\}$. □

The monotonicity property ensures that a union of two ECs yields no larger distance between $p_i$ and $q_i$ than its united parts. Hence, ECs violating $\beta$-likeness can be transformed to follow $\beta$-likeness by merge operations. The relative distance function instantiates basic $\beta$-likeness by the constraint $\mathsf{D}(p_i, q_i) = \frac{q_i - p_i}{p_i} \leq \beta$, where $p_i$ and $q_i$ are the distributions of any $\mathcal{SA}$ value $v_i \in \mathcal{V}$ in the whole table and an EC, respectively. This constraint amounts to an upper bound for the frequency of $v_i$ in any EC, $q_i$, namely $q_i \leq (1 + \beta) \cdot p_i$. Our *relative* distance function pays due attention to less frequent $\mathcal{SA}$ values. However, this function provides a meaningful frequency bound only if $(1+\beta) \cdot p_i < 1$; it then caters for $\mathcal{SA}$ values whose frequency in $\mathcal{DB}$ is $p_i < \frac{1}{1+\beta}$. In our effort to pay due attention to such less frequent values, we have discriminated against $\mathcal{SA}$ values of frequency larger than $\frac{1}{1+\beta}$. Such values can assume frequency 1 in an EC. Thus, an adversary identifying that a person's record is within such an EC can infer the $\mathcal{SA}$ value of that person with 100% confidence. The disclosure of such frequent $\mathcal{SA}$ values may pose a privacy threat. To address this limitation, we provide a stronger, *enhanced* definition of $\beta$-likeness.

DEFINITION 3 (ENHANCED $\beta$-LIKENESS). *For table $\mathcal{DB}$ with sensitive attribute $\mathcal{SA}$, let $\mathcal{V} = \{v_1, \ldots, v_m\}$ be the $\mathcal{SA}$ domain, and $\mathcal{P} = (p_1, \ldots, p_m)$ the overall $\mathcal{SA}$ distribution in $\mathcal{DB}$. An EC $\mathcal{G}$ with $\mathcal{SA}$ distribution $\mathcal{Q} = (q_1, \ldots, q_m)$ is said to satisfy enhanced $\beta$-likeness, if and only if $\forall q_i$, $\mathsf{D}(p_i, q_i) = \frac{q_i - p_i}{p_i} \leq \min\{\beta, -\ln p_i\}$, where $\beta > 0$ is a threshold and $\ln p_i$ is the natural logarithm of $p_i$.*

The inequality constraint in the above definition implies that $q_i \leq (1 + \min\{\beta, -\ln p_i\}) \cdot p_i$. We can then define the *upper bound* that enhanced $\beta$-likeness imposes on the frequency of $v_i$ in an EC by function $f(p_i) = (1 + \min\{\beta, -\ln p_i\}) \cdot p_i$, which can be decomposed as follows.

$$f(p_i) = \begin{cases} p_i(1+\beta), & 0 < p_i \leq e^{-\beta} \\ p_i(1-\ln p_i), & e^{-\beta} \leq p_i \leq 1 \end{cases} \quad (1)$$

The first segment of $f(p_i)$ is a linear, monotonically increasing function of $p_i$. The second segment is a concave, also monotonically increasing function of $p_i$, with derivative $-\ln p_i$. The two segments meet at $p_i = e^{-\beta}$. In effect, $f(p_i)$ is a continuous, monotonically increasing function of $p_i$ in $(0, 1]$ with $f(0) = 0$ and $f(1) = 1$. Intuitively, the second segment *bends* the function's slope so as not to exceed the maximum value of 1. The monotonicity of $f(p_i)$ implies that an EC $\mathcal{G}$ following the enhanced $\beta$-likeness constraint obeys the following properties:

1. The maximum frequency of an $\mathcal{SA}$ value $v_i$ in $\mathcal{G}$ is less than 1, i.e., $f(p_i) < 1$ for any $p_i < 1$.

2. For two $\mathcal{SA}$ values $v_i$ and $v_\ell$, such that $p_i < p_\ell$, the maximum allowed frequency of $v_i$ in $\mathcal{G}$ is less than that of $v_\ell$, i.e., $f(p_i) < f(p_\ell)$.

3. For an $\mathcal{SA}$ value $v_i$ that is 'infrequent' in table $\mathcal{DB}$, with $p_i \leq e^{-\beta}$, its frequency in $\mathcal{G}$ is at most $\beta$ times larger than $p_i$, i.e., $q_i \leq f(p_i) = (1+\beta) \cdot p_i$.

4. For an $\mathcal{SA}$ value $v_i$ that is 'frequent' in $\mathcal{DB}$, with $p_i > e^{-\beta}$, its frequency in $\mathcal{G}$ is at most $-\ln p_i$ times larger than $p_i$, i.e., $q_i \leq f(p_i) = (1 - \ln p_i) \cdot p_i < (1+\beta) \cdot p_i$.

These properties protect privacy for all $\mathcal{SA}$ values: infrequent values receive due attention, while more frequent ones are disallowed from assuming frequency values of 1. The $\beta$ parameter defines the privacy constraint for less frequent values, as well as the frequency threshold $e^{-\beta}$ above which the privacy constraint assumes a default form independent of $\beta$. This framework applies for any monotonic upper-bound function. Our choice of $\ln p_i$ is only a convenient choice that confers the desirable properties. As enhanced $\beta$-likeness provides more robust privacy than basic $\beta$-likeness, in the following we focus on it. Unless otherwise specified, henceforth by $\beta$-likeness we mean its enhanced form.

While (enhanced) $\beta$-likeness defines only an upper bound on $q_i$, the cognate $\delta$-disclosure-privacy model [3] amounts to two bounds on $q_i$, demanding that $\left|\log(\frac{q_i}{p_i})\right| < \delta$, or, equivalently, $e^{-\delta} \cdot p_i < q_i < e^{\delta} \cdot p_i$. Furthermore, there is a fundamental conceptual difference between $\beta$-likeness and $\delta$-disclosure-privacy: the former always disallows $q_i$ values equal to 0, and *can* allow $q_i$ values arbitrarily close to 1 (as its upper bound can assume values larger than 1), while the latter allows any $q_i$ value less than $p_i$, but always disallows $q_i$ values equal to 1 (its upper bound being strictly less than 1). We argue that both these choices are more reasonable than those made by $\delta$-disclosure-privacy. Moreover, we re-iterate that the introduction of $\delta$-disclosure-privacy in [3] was not accompanied by an anonymization algorithm tailored therefor; the model was only used as a tool to argue for a negative result, namely that existing $k$-anonymization algorithms [18], adapted to $\delta$-disclosure-privacy, yield unacceptably high information loss [3]. In contrast, our work aims at a positive result.

## 4. GENERALIZATION-BASED SCHEME

In this section we first introduce the metrics to measure the information loss by the generalization. Then we present an observation, which motivates our algorithm. After that, we design our generalization-based algorithm customized for $\beta$-likeness.

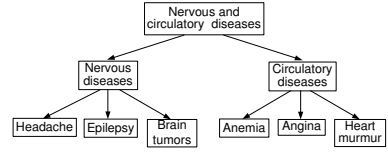

**Figure 1: Domain hierarchy for diseases**

### 4.1 Information Loss Metrics

To solve the problem posed by the $\beta$-likeness model, we need to fulfill the $\beta$ constraint while giving up little information. We use an *information loss metric* to assess the amount of information ceded for the sake of privacy. Different utility objectives would require different metrics. When the purpose the data is to be used for is not known in advance, a general metric can be used, as in [12].

Assume a set of $QI$ attributes $QI = \{A_1, \ldots, A_d\}$ and an EC $\mathcal{G}$. Given a numerical attribute $NA \in QI$, let $[L_{NA}, U_{NA}]$ be its domain and $[l_{NA}^{\mathcal{G}}, u_{NA}^{\mathcal{G}}]$ the (generalized) range of its values in $\mathcal{G}$; then the information loss (IL) regarding $NA$ in $\mathcal{G}$ is:

$$\mathcal{IL}_{NA}(\mathcal{G}) = \frac{u_{NA}^{\mathcal{G}} - l_{NA}^{\mathcal{G}}}{U_{NA} - L_{NA}} \quad (2)$$

Given a categorical attribute $CA$, we surmise a generalization hierarchy $\mathcal{H}_{CA}$ on its domain (Fig. 1). Let $a$ be the lowest common ancestor of all $CA$ values in $\mathcal{G}$; then, the IL regarding $CA$ in $\mathcal{G}$ is:

$$\mathcal{IL}_{CA}(\mathcal{G}) = \begin{cases} 0, & |\text{leaves}(a)| = 1 \\ \frac{|\text{leaves}(a)|}{|\text{leaves}(\mathcal{H}_{CA})|}, & \text{otherwise} \end{cases} \quad (3)$$

where leaves$(a)$ is the set of leaves under $a$, and leaves$(\mathcal{H}_{CA})$ the set of all leaves in $\mathcal{H}_{CA}$. Then the total IL of $\mathcal{G}$ is:

$$\mathcal{IL}(\mathcal{G}) = \sum_{i=1}^{d} w_i \times \mathcal{IL}_{A_i}(\mathcal{G}) \quad (4)$$

where $w_i$ is a weight for $A_i$, with $\sum_{i=1}^{d} w_i = 1$. In our experiments we set $w_i = \frac{1}{d}$. The Average Information Loss on a table $\mathcal{DB}$, published as a collection of ECs $S_{\mathcal{G}}$, is:

$$AIL(S_{\mathcal{G}}) = \frac{\sum_{\mathcal{G} \in S_{\mathcal{G}}} |\mathcal{G}| \times \mathcal{IL}(\mathcal{G})}{|\mathcal{DB}|} \quad (5)$$

We aim to attain $\beta$-likeness on $\mathcal{DB}$ at a low value of $AIL(S_{\mathcal{G}})$.

### 4.2 An Observation

The intuition behind our generalization-based method emanates from the following observation. Assume $\mathcal{DB}$ is partitioned into a set of buckets by a 'group-by' on $\mathcal{SA}$. If we form ECs by selecting from each bucket a number of tuples *proportional* to its size, then the $\mathcal{SA}$ distribution in the formed EC will be *the same* as the global distribution. On the other hand, if we partition $\mathcal{DB}$ into buckets allowing (all tuples of) more than one $\mathcal{SA}$ value per bucket, and then form ECs in a similar fashion, then there will be some variation in $\mathcal{SA}$ distributions among ECs. We aim to configure this process so as to allow for such variation to the extent permitted by the $\beta$ constraint. An akin methodology is followed in SABRE [4], an algorithm for the $t$-closeness model. Yet, unfortunately, SABRE *cannot* be applied on other distribution-based models, as it caters to the particular requirements of $t$-closeness, looking at the semantic distance between $\mathcal{SA}$ values in order to bound the EMD-difference of distributions between each EC and the overall table. In contrast, our algorithm should bound the variation in each $\mathcal{SA}$ value's frequency. The following two definitions clarify our intuition.

DEFINITION 4. *Given a table $\mathcal{DB}$ with sensitive attribute $\mathcal{SA}$, a set of buckets $\varphi$ forms an* exact bucket partition *of $\mathcal{DB}$ iff $\bigcup_{\forall \mathcal{B} \in \varphi} \mathcal{B} = \mathcal{DB}$, while each $\mathcal{SA}$ value (tuple) appears in* exactly one *bucket.*

DEFINITION 5 (PROPORTIONALITY CONDITION). *Let $\varphi$ be a bucket partition of $\mathcal{DB}$. Assume that an EC, $\mathcal{G}$, is formed with $x_j$ tuples from bucket $\mathcal{B}_j \in \varphi$, $j = 1, 2, \ldots, |\varphi|$. $\mathcal{G}$ abides to the* proportionality condition *with respect to $\varphi$, iff the values $x_j$ are proportional to $|\mathcal{B}_j|$, i.e., $x_1 : x_2 : \cdots : x_{|\varphi|} = |\mathcal{B}_1| : |\mathcal{B}_2| : \cdots : |\mathcal{B}_{|\varphi|}|$.*

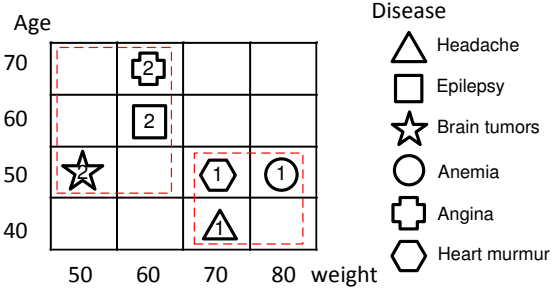

Figure 2: Improved information quality

EXAMPLE 1. *Consider Table 1, where {weight, age} is the QI, and disease is the $\mathcal{SA}$. The diagram in Figure 2 shows the QI-space and the distribution of tuples, with each QI attribute corresponding to a dimension. A bucket partition $\varphi$ of this table could consist of six buckets of one tuple each, with $\mathcal{SA}$ values* headache, epilepsy, brain tumors, anemia, angina, *and* heart murmur, *respectively. Taking one tuple from each of those, we could build a single EC satisfying 0-likeness. Still, such an EC covers the entire QI-space, incurring high information loss. An alternative bucket partition could consist of three two-tuple buckets, $\varphi = \{\mathcal{B}_1, \mathcal{B}_2, \mathcal{B}_3\}$, with* headache *and* epilepsy *in bucket $\mathcal{B}_1$,* brain tumors *and* anemia *in $\mathcal{B}_2$, and the rest in $\mathcal{B}_3$. We can then build two ECs, taking one tuple from each bucket, as shown in Figure 2. Tuples in the same EC are labeled by the same number in the figure. This partitioning achieves better information quality, as the areas of ECs in QI-space are smaller.*

While the bucket partition in the above example enables higher information quality, it no longer abides by 0-likeness. Still, it satisfies $\beta$-likeness, for $\beta \geq 1$, with respect to Table 1. In general, it suffices to create ECs so that they attain $\beta$-likeness for a given $\beta > 0$. We propose an algorithm that does so in two phases: it first partitions tuples into buckets, and then determines the number of tuples each EC needs to draw from each bucket.

## 4.3 Bucketization Phase

Let the $\mathcal{SA}$ domain be $\mathcal{V} = \{v_1, v_2, \ldots, v_m\}$ and the overall distribution of $\mathcal{SA}$ values $\mathcal{P} = (p_1, p_2, \ldots, p_m)$. We partition $\mathcal{V}$ into subsets, and use them to divide $\mathcal{DB}$ into a bucket partition $\varphi$; all tuples in $\mathcal{DB}$ with $\mathcal{SA}$ values in the same subset of $\mathcal{V}$ are pushed to a single bucket of $\varphi$. Assume EC $\mathcal{G}$ draws $x_j$ tuples from bucket $\mathcal{B}_j \in \varphi$, $j = 1, 2, \ldots, |\varphi|$, and let $V_j$ be the subset of $\mathcal{SA}$ values in $\mathcal{B}_j$. Then, in the worst case, *all* $x_j$ tuples may have the least frequent $\mathcal{SA}$ value in $V_j$, $v_{\ell_j}$, with $p_{\ell_j} = \min_{v_i \in V_j}\{p_i\}$, hence the frequency of $v_{\ell_j}$ in $\mathcal{G}$ will be $q_{\ell_j} = \frac{x_j}{|\mathcal{G}|}$; $\beta$-likeness should hold in this case too, i.e., it should be $\frac{x_j}{|\mathcal{G}|} \leq f(p_{\ell_j}) = (1+\min\{\beta, -\ln(p_{\ell_j})\}) \cdot p_{\ell_j}$, as the following theorem defines.

THEOREM 1 (ELIGIBILITY CONDITION). *Let $\varphi$ be a bucket partition of $\mathcal{DB}$ with sensitive attribute $\mathcal{SA}$, $\mathcal{G}$ an EC formed with $x_j$ tuples from bucket $\mathcal{B}_j \in \varphi$, $V_j$ the set of $\mathcal{SA}$ values in $\mathcal{B}_j$, and $p_{\ell_j} = \min_{v_i \in V_j}\{p_i\}$, $j = 1, 2, \ldots, |\varphi|$. If $\forall j \in \{1, 2, \ldots, |\varphi|\}$, $\frac{x_j}{|\mathcal{G}|} \leq f(p_{\ell_j})$, then $\mathcal{G}$ follows $\beta$-likeness.*

PROOF. For any $\mathcal{SA}$ value $v_k \in \mathcal{V}$, let $\mathcal{B}_j \in \varphi$ be the single bucket that contains tuples in $\mathcal{DB}$ with $v_k$ as their $\mathcal{SA}$ value, hence $v_k \in V_j$. Since $\mathcal{G}$ draws $x_j$ tuples from $\mathcal{B}_j$, the frequency of $v_k$ in $\mathcal{G}$ is $q_k \leq \frac{x_j}{|\mathcal{G}|} \leq f(p_{\ell_j}) \leq f(p_k)$. Expanding to all $v_k \in \mathcal{V}$, we conclude that $\mathcal{G}$ follows $\beta$-likeness. □

Theorem 1 defines the *eligibility condition* for an EC to follow $\beta$-likeness. However, it does not provide a way to specify a particular number of tuples $x_j$ to choose from a given bucket $\mathcal{B}_j$, i.e., it offers no guidance on how to construct a $\beta$-likeness-complying anonymization. To overcome this lack of guidance, we assume that ECs are formed following the *proportionality condition*. Under this assumption, it holds that $\frac{x_j}{|\mathcal{G}|} = \frac{|\mathcal{B}_j|}{|\mathcal{DB}|} = \sum_{v_i \in V_j} p_i$, and the next lemma can be easily deduced from Theorem 1.

LEMMA 2. *Let $\mathcal{G}$ be an EC that follows the proportionality condition with respect to a bucket partition $\varphi$ of $\mathcal{DB}$ with sensitive attribute $\mathcal{SA}$, $V_j$ the set of $\mathcal{SA}$ values in bucket $\mathcal{B}_j \in \varphi$, and $p_{\ell_j} = \min_{v_i \in V_j}\{p_i\}$, $j = 1, 2, \ldots, |\varphi|$. If $\forall j \in \{1, 2, \ldots, |\varphi|\}$, $\sum_{v_i \in V_j} p_i \leq f(p_{\ell_j})$, then $\mathcal{G}$ follows $\beta$-likeness.*

Lemma 2 defines the condition that the frequencies of a subset of $\mathcal{SA}$ values $V_j \subset \mathcal{V}$ should obey, so that, if all values in $V_j$ are put in the same bucket $\mathcal{B}_j$ by a bucket partition $\varphi$, then ECs obeying the proportionality condition with respect to $\varphi$ satisfy $\beta$-likeness. This condition is trivially satisfied by a strict partition having a *single* $\mathcal{SA}$ value per bucket. We aim at a *looser* bucket partition that satisfies the condition of Lemma 2 in a non-trivial manner, with *more than one* distinct $\mathcal{SA}$ values per bucket (as in Example 1).

---

**Function** DPpartition ($\mathcal{DB}, \mathcal{SA}$)

**1** Let $\mathcal{V} = \{v_1, v_2, \ldots, v_m\}$, $\mathcal{P} = (p_1, p_2, \ldots, p_m)$;
**2** Assume that $p_n \leq p_{n+1}$, where $n = 1, 2, \ldots, m - 1$;
**3** $N[0] = 0$;
**4** $S[0] = 0$;
**5** **for** *e=1 to m* **do**
**6**   $N[e] = N[e - 1] + 1$;
**7**   $S[e] = e$;
**8**   $b = e - 1$;
**9**   **while** $b > 0$ and Combinable$(b, e) = true$ **do**
**10**     **if** $N[b - 1] + 1 < N[e]$ **then**
**11**       $N[e] = N[b - 1] + 1$;
**12**       $S[e] = b$;
**13**     $b = b - 1$;
**14** Initialize $\varphi$ to be empty;
**15** $e = m$;
**16** **while** $e > 0$ **do**
**17**   $b = S[e]$;
**18**   Create bucket $\mathcal{B}$, having tuples with $\mathcal{SA}$ values in $\{v_b, v_{b+1}, \ldots, v_e\}$;
**19**   $\varphi = \varphi \cup \{\mathcal{B}\}$;
**20**   $e = S[e] - 1$;
**21** Return $\varphi$;

---

We develop a *bucketization* scheme for this task. We start out by representing, $\mathcal{P}$, the set of $\mathcal{SA}$ frequencies in $\mathcal{DB}$, in ascending order, $p_i \leq p_{i+1}$, $i = 1, \ldots, m - 1$. By Lemma 2, a set of consecutive $\mathcal{SA}$ values in $\mathcal{V}$, $v_b, v_{b+1}, \ldots, v_e$, are allowed to be in the same bucket provided that $\sum_{i=b}^{e} p_i < f(p_\ell)$, where $p_\ell = \min\{p_b, p_{b+1}, \ldots, p_e\}$. Our scheme, presented in Function DPpartition, partitions $\mathcal{V}$ by dynamic programming, so as to minimize the number of buckets. Let $N[e]$ denote the minimum number of buckets to which we can partition the prefix of $e$ elements in $\mathcal{V}$, i.e., $v_1, v_2, \ldots, v_e$. The value of $N[e]$ is calculated recursively as:

$$N[e] = \min_{\{b \mid \mathsf{Combinable}(b,e)=true\}} \{N[b-1]\} + 1 \qquad (6)$$

Function $\mathsf{Combinable}(b, e)$ checks whether $\mathcal{SA}$ values $v_b, \ldots, v_e$ can make a bucket, i.e., whether $\sum_{i=b}^{e} p_i < f(p_\ell)$, where $p_\ell = \min\{p_b, p_{b+1}, \ldots, p_e\}$. The DP base is $N[0] = 0$.

DPpartition has two parts. The first part (steps 3-13) runs the DP recursion of Equation 6 to evaluate the final minimum value $N[m]$ and split $\mathcal{V}$ into segments accordingly; thereby, it needs to assess the combinability of $m^2$ potential buckets. To assess combinability, we maintain the running $\sum p_i$ within a bucket, updated in $O(1)$ at each step; the $\min\{p_i\}$ within a bucket is its first element. The complexity of this part is $O(m^2)$. The second part (steps 14-20) uses the first-part results to build the bucket partition. Tuples with the $\mathcal{SA}$ values in the same segment make a bucket (step 18), in $O(|\mathcal{DB}|)$. The overall time complexity is $O(m^2 + |\mathcal{DB}|)$.

### 4.4 Reallocation Phase

The bucketization phase of our scheme delivers a bucket partition $\varphi$ of $\mathcal{DB}$. We have so far assumed that ECs are formed from $\varphi$ following the proportionality condition. However, a strict adherence to this condition may result in large ECs, incurring high information loss. For example, if the size of some bucket $\mathcal{B}_j \in \varphi$ is a prime number (other than 2), then, to strictly follow the proportionality condition, we should form an EC out of the whole table. We should rather relax the condition: it should suffice that the number of tuples $x_j$ chosen from bucket $\mathcal{B}_j$ in EC $\mathcal{G}$ be *approximately* proportional to the size of $\mathcal{B}_j$, i.e., $\frac{x_j}{|\mathcal{G}|} \approx \frac{|\mathcal{B}_j|}{|\mathcal{DB}|} = \sum_{v_i \in V_j} p_i$. The rationale for this relaxation is as follows: the bucket partition $\varphi$ returned by DPpartition obeys the inequality $\sum_{v_i \in V_j} p_i \leq f(p_{\ell_j})$ (Lemma 2). Then, if $\frac{x_j}{|\mathcal{G}|} \approx \sum_{v_i \in V_j} p_i$ (i.e., if we draw tuples into ECs approximately proportionally to the size of the bucket they hail from), then the *eligibility condition* $\frac{x_j}{|\mathcal{G}|} \leq f(p_{\ell_j})$ (Theorem 1), and therefore $\beta$-likeness, will be still easily achievable.

To ensure $\beta$-likeness, we determine the EC sizes by constructing a binary tree, the ECTree, in a top-down fashion. We start with a bucket partition $\varphi = \{\mathcal{B}_1, \ldots, \mathcal{B}_{|\varphi|}\}$. The root of the tree $r$ represents a potential EC that contains all tuples in $\mathcal{DB}$, i.e., $|\mathcal{B}_j|$ tuples from bucket $\mathcal{B}_j$. We denote these contents as $r = [|\mathcal{B}_1|, \ldots, |\mathcal{B}_{|\varphi|}|]$. This can be a valid EC, but we prefer smaller ones. Thus, we proceed to split $r$ into two children (each representing an EC), dividing each $\mathcal{B}_j$ into $\mathcal{B}_j^1$ and $\mathcal{B}_j^2$. The root's left child $c_L$ contains $\mathcal{B}_j^1$ and the right child $c_R$ contains $\mathcal{B}_j^2$, $j = 1, 2, \ldots, |\varphi|$. To ensure that $\mathcal{B}_j^1$ and $\mathcal{B}_j^2$ have *approximately* the same size, we set $|\mathcal{B}_j^1| = \mathrm{round}\left(\frac{|\mathcal{B}_j|}{2}\right)$ and $|\mathcal{B}_j^2| = |\mathcal{B}_j| - |\mathcal{B}_j^1|$. The split is allowed only if both $c_L$ and $c_R$ satisfy the eligibility condition (Theorem 1), hence can form ECs satisfying $\beta$-likeness. Assume the left child of $r$ is $c_L = [|\mathcal{B}_1^1|, \ldots, |\mathcal{B}_{|\varphi|}^1|]$. Then, for the eligibility condition to be satisfied, it should hold that $\frac{|\mathcal{B}_j^1|}{\sum_{n=1}^{|\varphi|} |\mathcal{B}_n^1|} \leq f(p_{\ell_j})$. An analogous condition applies for $c_R$. If splitting $r$ into $c_L$ and $c_R$ is allowed, we proceed to check whether we can split $c_L$ and $c_R$ themselves. When no node can be split further, we get a final ECTree, in which each leaf node configures the number of tuples an EC should get from each bucket. A simple function, biSplit$(\varphi)$, returns the list of leaf nodes. Example 2 illustrates this process.

EXAMPLE 2. *Let* disease *be a categorical $\mathcal{SA}$ with the domain hierarchy of Figure 1. Consider a table, containing 2 tuples with* headache, *3 with* epilepsy, *3 with* brain tumors, *3 with* anemia, *4 with* angina, *and 4 with* heart murmur. *Assume* $\beta = 2$. *The overall $\mathcal{SA}$ distribution is* $\mathcal{P} = (p_1, p_2, p_3, p_4, p_5, p_6) = (\frac{2}{19}, \frac{3}{19}, \frac{3}{19}, \frac{3}{19}, \frac{4}{19}, \frac{4}{19})$. $f(p_1) \approx 0.31$, $f(p_2) = f(p_3) = f(p_4) \approx 0.45$, *and* $f(p_5) = f(p_6) \approx 0.54$. *The bucketization phase returns a bucket partition of the table,* $\varphi = \{B_1, B_2, B_3\}$, *where $B_1$ accommodates tuples with $\mathcal{SA}$ values* headache *and* epilepsy, *$B_2$* brain tumors *and* anemia, *and $B_3$ the remaining two. The root node $r = [5, 6, 8]$ in Figure 3 represents an EC with 5 tuples from $B_1$, 6 from $B_2$, and 8 from $B_3$ (i.e., all tuples in the table). We split $r$ into $c_1 = [2, 3, 4]$ and $c_2 = [3, 3, 4]$. Then EC $c_1$ has size 9, and contains 2 tuples from $B_1$ with $\frac{2}{9} < \min\{f(p_1), f(p_2)\}$, 3 from $B_2$ with $\frac{3}{9} < \min\{f(p_3), f(p_4)\}$, and 4 from $B_3$ with $\frac{4}{9} < \min\{f(p_5), f(p_6)\}$. Thus, $c_1$ obeys the eligibility condition (Theorem 1). Likewise, $c_2$ also satisfies the condition. Thus, splitting $r$ into $c_1$ and $c_2$ is allowed. Recursively, we can split $c_1$ into $[1, 1, 2]$ and $[1, 2, 2]$. When we try to split $c_2$ into $g_1 = [1, 1, 2]$ and $g_2 = [2, 2, 2]$, we find that $g_2$ does not satisfy the eligibility condition, as $\frac{2}{6} > \min\{f(p_1), f(p_2)\}$, hence this split is not allowed. Figure 3 shows the final tree, with each leaf node indicating the number of tuples an EC should draw from each bucket. In the general case, an EC could also draw 0 tuples from some bucket.*

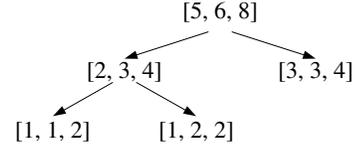

**Figure 3: Dynamic EC size determination**

### 4.5 BUREL

We now put the above phases together to devise BUREL, an algorithm that *BU*cketizes tuples into buckets and *RE*allocates them from buckets to ECs so as to attain $\beta$-*L*ikeness. The distinctive and novel feature of this algorithm, as opposed to algorithms for $k$-anonymity, $\ell$-diversity, and $t$-closeness, is that it distinguishes among $\mathcal{SA}$ values by their *frequencies* and builds its operation and reasoning around this frequency-based partitioning.

The *bucketization phase* of BUREL returns $\varphi$, a bucket partition of $\mathcal{DB}$ (step 2). Then, its *reallocation phase* (function biSplit) determines the number of tuples each EC should draw from each bucket at a leaf of the ECTree and returns a list of arrays $S_a$ containing these size values (step 3). Specific ECs following the prescribed sizes are then materialized (steps 4-9). Given an array $a \in S_a$, BUREL retrieves $a_j$ tuples from $\mathcal{B}_j \in \varphi$, where $a_j$ is the $j^{th}$ element of $a$ and $j = 1, 2, \ldots, |\varphi|$, and forms an EC $\mathcal{G}$ out of them (steps 6-8).

| **Algorithm:** BUREL ( $\mathcal{DB}$, $\mathcal{SA}$, $\beta$ ) |
|---|
| **1** Let $\{v_1, v_2, \ldots, v_m\}$ be all the $\mathcal{SA}$ values in $\mathcal{DB}$, and $\{p_1, p_2, \ldots, p_m\}$ be their distributions; |
| **2** $\varphi = \mathsf{DPpartition}(\mathcal{DB}, \mathcal{SA})$; |
| **3** $S_a = \mathsf{biSplit}(\varphi)$; |
| **4 foreach** *array $a$ in $S_a$* **do** |
| **5**      Create an empty EC, say $\mathcal{G}$; |
| **6**      **foreach** $a_j$, $j^{th}$ *element of $a$* **do** |
| **7**          $ec_j = \mathsf{Retrieve}(\mathcal{B}_j, a_j)$; |
| **8**          add $ec_j$ to $\mathcal{G}$; |
| **9**      Output $\mathcal{G}$; |

When retrieving tuples from buckets, BUREL is indifferent to their $\mathcal{SA}$ values. The $\beta$-likeness between a constructed EC $\mathcal{G}$ and the whole table $\mathcal{DB}$ is guaranteed by Theorem 1; tuple selection is guided by information loss considerations, as prescribed by our information loss metric (Section 4.1). This metric requires the Minimum Bounding Boxes of ECs to be small. Accordingly, we employ function Retrieve($\mathcal{B}_i, a_i$) (step 7), which greedily picks tuples of similar $QI$ values. This greedy selection of nearby tuples works as follows: We define a multidimensional space with each $QI$ attribute as a dimension. The mapping to such a $QI$-space for a numerical $QI$ attribute $NA$ is straightforward. The axis of a categorical $QI$ attribute $CA$ is formed by the order provided by a pre-order traversal of the leaves in its domain hierarchy $\mathcal{H}_{CA}$. Each tuple is represented as a point in this $QI$-space. When forming an EC $\mathcal{G}$, BUREL first randomly picks a tuple $x$ from a bucket of $\varphi$ in $\mathcal{G}$, and then finds the $a_j$ nearest neighbors (by Euclidean distance) of $x$ in each bucket $\mathcal{B}_j$, $j = 1, 2, \ldots, |\varphi|$, and adds them into $\mathcal{G}$, until the size specifications are satisfied. Still, this process can be computationally demanding even with an index structure [8]. Thus, we devise an efficient heuristic using the *Hilbert curve* [24], a continuous fractal that maps regions of $QI$-space, hence tuples, to 1D Hilbert values. Tuples close in $QI$-space are *likely* to have nearby Hilbert values. BUREL sorts tuples in $\mathcal{B}_j$ by their Hilbert values, and uses this order to select the $a_j$ nearest neighbors of a tuple $x$ within *each* bucket. We find the nearest Hilbert-neighbor $\bar{x}$ of $x$ within bucket $\mathcal{B}_j$ by binary search, and then expand to the next closest $a_j$ neighbors to $x$. The average time complexity for this search operation is $O\big(|S_\mathcal{G}||\varphi|\cdot(\log\frac{|\mathcal{DB}|}{|\varphi|}+\frac{|\mathcal{DB}|}{|S_\mathcal{G}||\varphi|})\big)$, where $|\varphi|$ is the number of buckets, $\frac{|\mathcal{DB}|}{|\varphi|}$ the average size of a bucket, and $\frac{|\mathcal{DB}|}{|S_\mathcal{G}||\varphi|}$ the average number of tuples drawn from a bucket to form an EC.

## 5. PERTURBATION-BASED SCHEME

Our generalization-based solution, achieves $\beta$-likeness and also provides identity anonymity, like all generalization-based methods do. However, in case a data set contains a few remote outliers, these outliers may force a highly unsatisfactory solution by generalization. Similarly unsatisfactory solutions can be obtained in case of extremely infrequent $\mathcal{SA}$ values. For example, consider a dataset $\mathcal{DB}$, in which only one tuple $t$ has $\mathcal{SA}$ value $v$. Then, to attain 1-likeness, we would have to create an EC containing $t$ and at least half of the tuples in $\mathcal{DB}$. We deduce that an alternative solution is desirable in order to handle such irregular cases, even at the expense of identity anonymity. To that end, we propose an approach that anonymizes each tuple independently by perturbing $\mathcal{SA}$ values while preserving $QI$ values intact. We reiterate that, for a given $\mathcal{SA}$ value $v_i$, $\beta$-likeness considers its frequency in the whole table, $p_i$, as prior confidence, and constrains an adversary's information gain on $v_i$ after seeing the published data, bounding the posterior confidence. We aim to achieve this target by perturbing $\mathcal{SA}$ values only; our scheme resembles a randomized response procedure, albeit having a different perturbation probability for each $\mathcal{SA}$ value.

DEFINITION 6 ($\beta$-LIKENESS BY PERTURBATION). *Given a table $\mathcal{DB}$ with sensitive attribute $\mathcal{SA}$, let $\mathcal{V} = \{v_1, \ldots, v_m\}$ be the $\mathcal{SA}$ domain, and $\mathcal{P} = (p_1, \ldots, p_m)$ the overall $\mathcal{SA}$ distribution in $\mathcal{DB}$. A perturbation on $\mathcal{DB}$ that randomizes $\mathcal{SA}$ values satisfies $\beta$-likeness, iff the adversarial posterior confidence in $v_i \in \mathcal{V}$ after seeing the randomized data is at most $f(p_i)$, $i = 1, 2, \ldots, m$.*

To build a solution that achieves $\beta$-likeness by perturbation, we adapt the concept of *upward* ($\rho_1, \rho_2$)-privacy [10] as follows.

DEFINITION 7 (($\rho_{1i}, \rho_{2i}$)-PRIVACY). *Let $v_i \in \mathcal{V}$ be an original $\mathcal{SA}$ value, and $v \in \mathcal{V}$ be any $\mathcal{SA}$ value after perturbation. We say that ($\rho_{1i}, \rho_{2i}$)-privacy is satisfied on $v_i$, iff the adversarial prior confidence in $v_i$ is $\mathsf{C}(U = v_i) = \rho_{1i}$, and the posterior confidence after seeing $v$ is $\mathsf{C}(U = v_i | V = v) \leq \rho_{2i}$.*

While ($\rho_1, \rho_2$)-privacy does not distinguish among $\mathcal{SA}$ values, our adaptation does. Given these definitions, we can achieve $\beta$-likeness by setting $\rho_{1i} = p_i$ and $\rho_{2i} = f(p_i)$ for each $v_i \in \mathcal{V}$.

THEOREM 2. *Let $v_i \in \mathcal{V}$ be an original $\mathcal{SA}$ value, and $v \in \mathcal{V}$ be an $\mathcal{SA}$ value after perturbation, such that $\exists u \in \mathcal{V}$: $\Pr(u \to v) > 0$, where $u \to v$ denotes that $u$ has been perturbed to $v$. If it holds that*

$$\forall v_j \in \mathcal{V}: \frac{\Pr(v_i \to v)}{\Pr(v_j \to v)} \leq \frac{\rho_{2i}}{\rho_{1i}} \cdot \frac{1 - \rho_{1i}}{1 - \rho_{2i}} = \gamma_i \quad (7)$$

*then ($\rho_{1i}, \rho_{2i}$)-privacy is satisfied with $\rho_{1i} = \mathsf{C}(U = v_i) > 0$.*

PROOF. Assume that ($\rho_{1i}, \rho_{2i}$)-privacy is *not* satisfied, that is, $\mathsf{C}(U = v_i | V = v) > \rho_{2i}$. For the event of seeing $v$ it holds that $\mathsf{C}(V = v) = \sum_{\forall u \in \mathcal{V}} \mathsf{C}(U = u) \cdot \Pr(u \to v) > 0$, as $v$ must have been produced by some original value $u$. Let $v_j$ be an $\mathcal{SA}$ value least likely to have been perturbed to $v$, i.e.:

$$v_j \in \{u \in \mathcal{V} | \Pr(u \to v) = \min_{u' \in \mathcal{V}} \Pr(u' \to v)\}$$

By the definition of conditional probability it holds that:

$$\mathsf{C}(U = v_i | V = v) = \frac{\mathsf{C}(U = v_i) \cdot \Pr(v_i \to v)}{\mathsf{C}(V = v)} \quad (8)$$

and, since $v_j$ is least likely to have yielded $v$, it is:

$$\mathsf{C}(U \neq v_i | V = v) \geq \frac{\mathsf{C}(U \neq v_i) \cdot \Pr(v_j \to v)}{\mathsf{C}(V = v)} \quad (9)$$

Since, by our assumption, $\mathsf{C}(U = v_i | V = v) > \rho_{2i} > 0$ and $\mathsf{C}(U = v_i) = \rho_{1i} > 0$, from Eqs. (8) and (9) we get:

$$\frac{\mathsf{C}(U \neq v_i | V = v)}{\mathsf{C}(U = v_i | V = v)} \geq \frac{\Pr(v_j \to v)}{\Pr(v_i \to v)} \cdot \frac{\mathsf{C}(U \neq v_i)}{\mathsf{C}(U = v_i)} \quad (10)$$

Inequality (7) holds for $v_j$, thus we can rewrite Inequality (10) as:

$$\frac{1 - \mathsf{C}(U = v_i | V = v)}{\mathsf{C}(U = v_i | V = v)} \geq \frac{1}{\gamma_i} \cdot \frac{1 - \mathsf{C}(U = v_i)}{\mathsf{C}(U = v_i)} \quad (11)$$

Still, $\frac{1 - \mathsf{C}(U = v_i)}{\mathsf{C}(U = v_i)} = \frac{1 - \rho_{1i}}{\rho_{1i}}$, hence Inequality (11) yields:

$$\frac{1 - \mathsf{C}(U = v_i | V = v)}{\mathsf{C}(U = v_i | V = v)} \geq \frac{1}{\gamma_i} \cdot \frac{1 - \rho_{1i}}{\rho_{1i}} = \frac{1 - \rho_{2i}}{\rho_{2i}} \Rightarrow$$
$$\mathsf{C}(U = v_i | V = v) \leq \rho_{2i}$$

which contradicts our assumption. □

Due to Theorem 2, Inequality (7) provides a sufficient condition for $\beta$-likeness to hold. We aim to achieve this condition by uniform perturbation, which maximizes the utility of randomized data [2]. Given an input $\mathcal{SA}$ value $v_i \in \mathcal{V}$, uniform perturbation tosses a coin with probability $\alpha_i \in (0, 1]$ for *heads* and $1 - \alpha_i$ for *tails*, and, in the latter case, replaces $v_i$ by a randomly selected value $v \in \mathcal{V}$. Then:

$$\Pr(v_i \to v) = \begin{cases} \alpha_i + (1 - \alpha_i)/m & \text{if } v_i = v \\ (1 - \alpha_i)/m & \text{if } v_i \neq v \end{cases} \quad (12)$$

LEMMA 3. *Given any perturbed value $v$, $\Pr(v_i \to v)$ is maximized when $v_i = v$.*

PROOF. By Equation 12, if $v_i = v$, then $\Pr(v_i \to v) = \alpha_i + (1 - \alpha_i)/m$. For $v_j \neq v$, $\Pr(v_j \to v) = \frac{1 - \alpha_j}{m}$. Since $a_i, a_j \in (0, 1]$ it is $\Pr(v_i \to v) - \Pr(v_j \to v) = \frac{(m-1) \cdot \alpha_i + \alpha_j}{m} > 0$. □

For the sake of utility, we need to maximize the probability that input $\mathcal{SA}$ values remain unchanged, i.e., to set $\alpha_i$ as high as possible for each $v_i \in \mathcal{V}$. However, for a given $v_i$, the value of $\alpha_i$ should allow Inequality (7) to hold for $v = v_i$ and for any $v_j \neq v$; if it holds in these extreme cases, then it also holds for all other values of variables $v$ and $v_j$. Substituting the values given by Equation 12 for $v = v_i$ and $v_j \neq v$ in Inequality (7), we get $\frac{\alpha_i + (1-\alpha_i)/m}{(1-\alpha_j)/m} \leq \gamma_i$ $\forall j \in \{1, 2, \ldots, m\} \setminus \{i\}$. The worst-case value the denominator in the last inequality (i.e., the probability $\Pr(v_j \to v)$ of perturbing $v_j \neq v$ to $v$) can assume is $\mathsf{C}_M = \min_{h=1}^{m}\left\{\frac{1-\alpha_h}{m}\right\}$. To calculate a bound for $\alpha_i$, we require that the inequality holds in the worst case:

$$\frac{\alpha_i + (1-\alpha_i)/m}{\mathsf{C}_M} \leq \gamma_i \Leftrightarrow \quad (13)$$

$$\alpha_i \leq \frac{m \cdot \gamma_i \cdot \mathsf{C}_M - 1}{m-1} \quad (14)$$

Since by definition $\mathsf{C}_M \leq \frac{1-\alpha_i}{m}$, if Inequality (13) holds, then, for a given $i$, it will hold that $\frac{\alpha_i + (1-\alpha_i)/m}{(1-\alpha_i)/m} \leq \gamma_i$; consequently, it will be $\alpha_i \leq \frac{\gamma_i - 1}{\gamma_i + m - 1}$. Using this upper bound of $\alpha_i$, we infer that, for each $i \in \{1, 2, \ldots, m\}$ it will be $\frac{1-\alpha_i}{m} \geq \frac{1}{\gamma_i + m - 1}$. In effect, it should also be $\mathsf{C}_M = \min_{h=1}^{m}\left\{\frac{1-\alpha_h}{m}\right\} \geq \min_{h=1}^{m}\left\{\frac{1}{\gamma_h + m - 1}\right\} = \frac{1}{\gamma_\ell + m - 1}$, where $\gamma_\ell = \max_{h=1}^{m}\{\gamma_h\}$. We have thus derived a lower bound for $\Pr(v_j \to v)$ with $v_j \neq v$, namely $\mathsf{C}_M^L = \frac{1}{\gamma_\ell + m - 1}$. To ensure that Inequality (14) always holds, we must guarantee that it also holds for the lower-bound case; thus, the highest value we can safely assign to each $\alpha_i$ is $\alpha_i = \frac{m \cdot \gamma_i \cdot \mathsf{C}_M^L - 1}{m-1}$. Eventually:

THEOREM 3. *Perturbation by Eq. (12) with* $\alpha_i = \frac{m \cdot \gamma_i \cdot \mathsf{C}_M^L - 1}{m-1}$, $\gamma_i = \frac{\rho_{2i}}{\rho_{1i}} \cdot \frac{1 - \rho_{1i}}{1 - \rho_{2i}}$, $\rho_{1i} = p_i$, $\rho_{2i} = f(p_i)$, $\forall v_i \in \mathcal{V}$, *satisfies $\beta$-likeness.*

PROOF. Due to Lemma 3, for any $\mathcal{SA}$ value $v_j$, perturbation by Eq. (12) gives $\frac{\Pr(v_i \to v)}{\Pr(v_j \to v)} \leq \frac{\alpha_i + (1-\alpha_i)/m}{\mathsf{C}_M^L} = \gamma_i$, $i = 1, 2, \ldots, m$. Then, by Theorem 2, $(\rho_{1i}, \rho_{2i})$-privacy holds $\forall v_i \in \mathcal{V}$, hence $\beta$-likeness is satisfied. □

We now discuss how we reconstruct the original $\mathcal{SA}$ distribution from the perturbed data to answer aggregation queries, which are a basis of data mining tasks, as the following [33]:

```
SELECT COUNT(*) FROM Anonymized-data
WHERE pred(A_1) AND ... AND pred(A_λ)
AND pred(SA)
```

This query has predicates on $\lambda$ randomly selected $QI$ attributes and the $\mathcal{SA}$. For each of these $\lambda + 1$ attributes $A$, $pred(A)$ has the form of $A \in R_A$, where $R_A$ is an arbitrary interval in the domain of $A$.

Perturbation does not affect $QI$ values. We reconstruct the query's result by estimating the count of original $\mathcal{SA}$ values among those tuples that satisfy the query's $QI$ predicates, given the observed $\mathcal{SA}$ values. In particular, given an aggregation query, suppose that $\mathcal{S}_t$ is the set of tuples satisfying the predicates associated with $QI$ attributes, $A_1 \in R_{A_1}$ AND … AND $A_\lambda \in R_{A_\lambda}$. Let $\mathcal{S}'_t$ be the perturbed form of $\mathcal{S}_t$. Since perturbation randomizes only $\mathcal{SA}$ values, each tuple in $\mathcal{S}_t$ is perturbed in $\mathcal{S}'_t$ with its $QI$ value unchanged. Let $n_i$ be the number of tuples with $\mathcal{SA}$ value $v_i$ in $\mathcal{S}_t$, $i = 1, 2, \ldots, m$, and $e_i = \sum_{j=1}^{m} \Pr(v_j \to v_i) \cdot n_j$ the expected number of instances of $v_i$ in $\mathcal{S}'_t$. According to our previous discussion, if $j = i$, then $\Pr(v_j \to v_i) = \gamma_i \cdot \mathsf{C}_M^L$, else $\Pr(v_j \to v_i) = \frac{1 - \gamma_j \cdot \mathsf{C}_M^L}{m-1}$. Using the notation $X_i = \gamma_i \cdot \mathsf{C}_M^L$ and $Y_j = \frac{1 - \gamma_j \cdot \mathsf{C}_M^L}{m-1}$, we have $E = PM \times N$, where $E = <e_1, e_2, \ldots, e_m>$, $N = <n_1, n_2, \ldots, n_m>$, and

$$PM = \begin{bmatrix} X_1 & Y_2 & \ldots & Y_m \\ Y_1 & X_2 & \ldots & Y_m \\ \vdots & \vdots & \ddots & \vdots \\ Y_1 & Y_2 & \ldots & X_m \end{bmatrix}$$

A data recipient knows neither $E$ nor $N$, but only observes $E' = <e'_1, e'_2, \ldots, e'_m>$, where $e'_i$ is the number of occurrences of $\mathcal{SA}$ value $v_i$ in $\mathcal{S}'_t$. Thus, one can approximately reconstruct $N$ as $N' = PM^{-1} \times E' = <n'_1, n'_2, \ldots, n'_m>$, and estimate the answer to a given query as $est = \sum_{v_i \in R_{SA}} n'_i$, where $R_{SA}$ is the query interval of $pred(\mathcal{SA})$. To facilitate this reconstruction process, we publish the perturbed data along with matrix $PM$; we can also release the original global $\mathcal{SA}$ distribution $\mathcal{P}$ in order to render the publication model comparable to that offered by generalization.

## 6. EXPERIMENTAL EVALUATION

In this section we evaluate our schemes. Our prototypes were implemented in Java and the experiments ran on a Core2 Duo 2.33GHz CPU machine with 4GB RAM running Windows XP. We use the CENSUS dataset [1], which contains 500,000 tuples on 6 attributes as shown in Table 3. For categorical attributes, the value following the type is the height of the corresponding attribute hierarchy; for instance, attribute *marital status* is categorical and has a hierarchy of height 2. The first 5 attributes are potential $QI$-attributes; the last (*salary class*) is the $\mathcal{SA}$. By default, we take the first three attributes as $QI$. The least frequent $\mathcal{SA}$ value is 49, with frequency 0.2018%; the most frequent $\mathcal{SA}$ value is 12, with frequency 4.8402%; $\beta = 1$ produces frequency threshold $e^{-\beta} \approx 37\%$, which marks all $\mathcal{SA}$ values as 'infrequent', and allows the frequency of any $\mathcal{SA}$ value in any EC to be at most $4.8402\% \times 2 = 9.7\%$. Thus, 1 is a small $\beta$ value. We use $\beta \in \{1, 2, 3, 4, 5\}$. We generate 5 microdata tables by randomly picking 100K to 500K tuples from the dataset; the one of 500K tuples is our default dataset.

| Attribute | Cardinality | Type |
|---|---|---|
| Age | 79 | numerical |
| Gender | 2 | categorical (1) |
| Education Level | 17 | numerical |
| Marital Status | 6 | categorical (2) |
| Work Class | 10 | categorical (3) |
| Salary Class | 50 | sensitive attribute |

**Table 3: The CENSUS dataset**

We set the likeness threshold $\beta$ by default to 4. Then, given the application of *enhanced* $\beta$-likeness for any $\mathcal{SA}$ value $v_i$, if $p_i \leq e^{-4} = 0.018$, its frequency $q_i$ in any EC should not exceed $5p_i$; if $p_i > 1.8\%$, then it should be $q_i \leq (1 - \ln(p_i)) \cdot p_i$. We reiterate that these bounds apply to *each* $\mathcal{SA}$ value, while their definition accommodates both low-frequency and high-frequency values. The highest $\mathcal{SA}$ value frequency in our data set does not exceed 5%, so the frequency of any salary class in any EC will not exceed 20%.

### 6.1 Face-to-face with $t$-closeness

Our first task is to compare our new $\beta$-likeness privacy model to the predecessor distribution-based model of $t$-closeness. We argue that $\beta$-likeness provides a more informative and comprehensible privacy guarantee than $t$-closeness does. Still, in order to create an even playing field on which to compare $\beta$-likeness to $t$-closeness, we conducted three face-to-face comparisons as follows.

In the first comparison, for a given dataset $\mathcal{DB}$ and $\beta$, we let BUREL transform $\mathcal{DB}$ to $\mathcal{DB}_\beta$, satisfying $\beta$-likeness. We then measure the closeness $t_\beta$, by the $t$-closeness model, between $\mathcal{DB}_\beta$ and $\mathcal{DB}$, i.e., the maximum EMD of the $\mathcal{SA}$ distribution in an EC of $\mathcal{DB}_\beta$ from its distribution in $\mathcal{DB}$. We then apply $t$-closeness schemes tMondrian [20] and SABRE [4] on $\mathcal{DB}$ as well, with $t_\beta$ as the $t$-closeness threshold, to produce $\mathcal{DB}_{t_\beta}^M$ and $\mathcal{DB}_{t_\beta}^S$, respectively. Then $\mathcal{DB}_\beta$, $\mathcal{DB}_{t_\beta}^M$, and $\mathcal{DB}_{t_\beta}^S$ achieve the *same* privacy under the criterion of $t$-closeness, as expressed by $t_\beta$. Then we measure the $\beta$ value achieved by $\mathcal{DB}_{t_\beta}^M$ and $\mathcal{DB}_{t_\beta}^S$ with respect to $\mathcal{DB}$. Given that all three schemes achieve the same privacy in terms of $t$-closeness, we are interested to compare the privacy they achieve in terms of $\beta$-likeness. Figure 4(a) shows the results (in logarithmic y-axes), as a function of the given $\beta$ parameter. While all the three schemes are tuned to ensure the same $t$-closeness guarantee, BUREL provides consistently higher privacy by the criterion of $\beta$-likeness than SABRE and tMondrian. This result is expected, since $t$-closeness restricts only the cumulative difference between $\mathcal{SA}$ distributions, indifferent to the relative frequency difference of each individual $\mathcal{SA}$ value between an EC and the whole table.

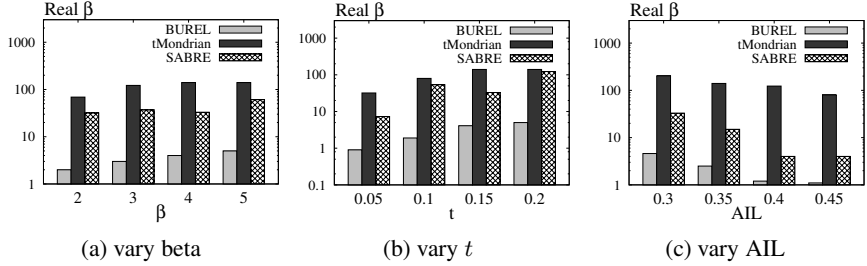

**Figure 4: Comparison to $t$-closeness**

Next, for a given dataset $\mathcal{DB}$ and closeness constraint $t$, we let tMondrian (SABRE) transform $\mathcal{DB}$ to $\mathcal{DB}^M_t$ ($\mathcal{DB}^S_t$), attaining $t$-closeness. We then let BUREL find, by binary search, a value $\beta_t$, such that, when it enforces $\beta_t$-likeness on $\mathcal{DB}$, it produces an anonymization $\mathcal{DB}_{\beta_t}$ characterized by the same (or smaller) closeness parameter $t$ as $\mathcal{DB}^M_t$ ($\mathcal{DB}^S_t$). Again we get three anonymized versions of $\mathcal{DB}$ that achieve the same privacy under $t$-closeness. While in our first comparison we arrived at this state starting out with a $\beta$ parameter, now we start out with a $t$ parameter. Thus, we avoid bias against $t$-closeness schemes. We now compare the $\beta$-likeness achieved by $\mathcal{DB}^M_t$ ($\mathcal{DB}^S_t$) to that of $\mathcal{DB}_{\beta_t}$, as a function of $t$. The results, shown in Figure 4(b), reaffirm our previous findings.

In our last experiment, given an AIL value $l$, we let BUREL determine, by binary search on its $\beta$ threshold, a value $\beta_l$, such that the data set $\mathcal{DB}_{\beta_l}$ it generates from $\mathcal{DB}$ with $\beta_l$ as the likeness threshold achieves AIL equal to (or smaller than) $l$. Likewise, we determine, by binary search, a value $t^M_l$ ($t^S_l$), which, used as the closeness threshold in tMondrian (SABRE), generates data set $\mathcal{DB}_{t^M_l}$ ($\mathcal{DB}_{t^S_l}$) with AIL near $l$ too, allowing for a small difference $\epsilon$. Thus, we obtain three data sets $\mathcal{DB}_{\beta_l}$, $\mathcal{DB}_{t^M_l}$, and $\mathcal{DB}_{t^S_l}$, generated by BUREL, tMondrian, and SABRE, respectively, which all have information loss near $l$; to ensure the comparison is not biased in favor of BUREL, we ensure its AIL value is *not greater* than those of the other algorithms. We then compare the privacy they achieve in terms of $\beta$-likeness. Figure 4(c) shows the results. Not surprisingly, BUREL provides the highest privacy again, followed by SABRE and tMondrian.

Our results testify that, other factors being equal, state-of-the-art $t$-closeness schemes fail by a wide margin (as indicated by the logarithmic y-axes) to achieve privacy good in terms of $\beta$-likeness. Thus, they reaffirm that $\beta$-likeness raises substantially different requirements from $t$-closeness, and requires a different approach.

## 6.2 Evaluation on Generalization

In this section we evaluate the performance of BUREL as a $\beta$-likeness algorithm in its own field. As there is no previous work on $\beta$-likeness, we employ two comparison benchmarks adopting some suggestions of related work. First, we devise an algorithm for $\beta$-likeness, following the conventional wisdom on designing algorithms for new privacy models: We adapt Mondrian [18], a $k$-anonymization algorithm, to the purposes of $\beta$-likeness, as previous works have done for other privacy models [22, 20, 3, 21].

Our adaptation, LMondrian, splits an EC only if both resultant ECs satisfy $\beta$-likeness. Second, we use the similar adaptation of Mondrian to $\delta$-disclosure-privacy suggested in [3], DMondrian. To render DMondrian comparable to BUREL and LMondrian, we set the value of $\delta$ so that the data anonymized by DMondrian obey $\beta$-likeness. As we have discussed, while $\beta$-likeness demands that an $\mathcal{SA}$ value's distribution in an EC be $q_i \leq (1+\min\{\beta, -\ln p_i\}) \cdot p_i$, for a given $\beta$, $\delta$-disclosure-privacy requires that $e^{-\delta} \cdot p_i < q_i < e^{\delta} \cdot p_i$, where $p_i$ is the overall distribution of $v_i$ in the whole dataset. Thus, an algorithm for $\delta$-disclosure-privacy achieves $\beta$-likeness for $\delta \leq \log(1+\min\{\beta, -\ln p_i\})$, for all $p_i$; in view of all $\mathcal{SA}$ values in $\mathcal{V}$, we set $\delta = \log\left(1+\min\left\{\beta, -\ln\left(\max_{v_i \in \mathcal{V}}\{p_i\}\right)\right\}\right)$. We first compare the three schemes with respect to average information loss and wall-clock time.

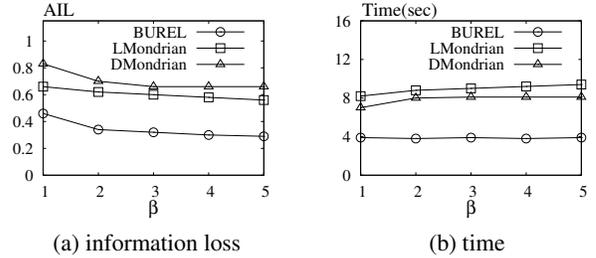

(a) information loss      (b) time

**Figure 5: Effect of varying $\beta$**

First, we study performance as a function of the $\beta$ threshold. Figure 5 shows the results. As $\beta$ grows, the constraint on the relative difference of each $\mathcal{SA}$ (i.e., *salary*) value frequency between an EC and the overall table is relaxed, hence information quality rises (Figure 5(a)). BUREL outperforms both LMondrian and DMondrian in information quality, showing the benefit of a scheme tailored for $\beta$-likeness. This result reconfirms the finding of [3] that a $k$-anonymization algorithm, adapted to $\delta$-disclosure-privacy, yields unacceptably high information loss; as we discussed, we aim at a *positive* result and propose a *better* alternative. In addition, given that $\delta$-disclosure-privacy overprotects data by imposing a constraint on *negative* information gain, LMondrian performs better than its stricter sibling, DMondrian. Remarkably, BUREL also outpaces both Mondrian-based schemes in efficiency (Figure 5(b)). Overall, BUREL achieves almost half the information loss of its Mondrian-based competitors in about half the time.

Next, we investigate the effect of $QI$ dimensionality (size), varying it from 1 to 5. As $QI$ dimensionality increases, the data become more sparse in $QI$ space, as more high-dimensional degrees of freedom are offered; thus, the formed ECs are more likely to have large minimum bounding boxes, and information quality degrades, as Figure 6(a) shows. The information loss of BUREL is again lower than that of the Mondrian-based methods. In addition, BUREL is again the fastest of the three (Figure 6(b)).

Our next experiment studies the effect of database size, varying the size of the microdata table from 100K to 500K tuples. Figure

1396

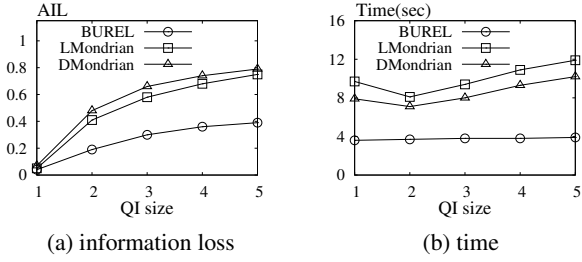

(a) information loss  (b) time

**Figure 6: Effect of varying $QI$**

7 presents our results. Interestingly, data size has no clear effect on information quality. This is due to the fact that, as the amount of tuples grows, more sensitive values are revealed, imposing their own requirements. The mere increase of data density does not help, as it would with simpler models like $k$-anonymity. Still, the elapsed time increases as the table size grows; BUREL is again found to be superior in both respects.

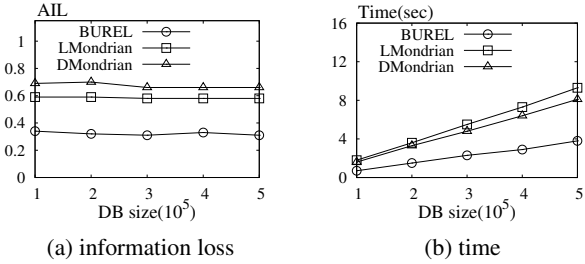

(a) information loss  (b) time

**Figure 7: Effect of varying dataset**

We now examine the utility of the generalized table by aggregation queries introduced in Section 5. Each predicate $pred(A)$ in the query has the form of $A \in R_A$. Let *expected selectivity* over the table be $0 < \theta < 1$. Assuming data are uniformly distributed, $\theta$ can be achieved if each attribute $A$ selects records within a range of length $|A| \cdot \theta_A$ of its domain, such that $(\theta_A)^{\lambda+1} = \theta$. In effect, the length of $R_A$ should be $|A| \cdot \theta^{\frac{1}{\lambda+1}}$, where $|A|$ is the domain length of attribute $A$. Given a query, the precise result $prec$ is computed from the original table, and an estimated result $est$ is obtained from the anonymized table. To calculate $est$, we assume that tuples in each EC are uniformly distributed, and consider the intersection between the query and the EC. We define $\frac{|est-prec|}{prec} \times 100\%$ as the *relative error*. We measure the *median relative error* in a workload of 10K queries. Relative error is undefined when $prec$ is 0. If $prec$ in a query is 0, we drop that query.

In our first experiment, we use the first 5 attributes in Table 3 as $QI$, with expected selectivity $\theta = 0.1$, and vary the dimensionality of the query, i.e. the number of $QI$ attributes $\lambda$ on which predicates are defined. As these attributes contribute to the error, the increase of $\lambda$ exercises a negative effect on error. However, as $\lambda$ grows, the length of the query range $R_A$ in the domain of each queried attribute also grows (for constant $\theta$); thereby, the minimum bounding box of an EC becomes more likely to be entirely contained in the query region. In effect, the error does not depend monotonically on $\lambda$ (Figure 8(a)); it does not matter much how many attributes a given selectivity $\theta$ is shared among. In the next experiment, we fix $\lambda$ to 3, $\theta$ to 0.1, and vary $\beta$. Figure 8(b) shows the results. As $\beta$ grows, the privacy requirement is relaxed, hence information quality rises and the error drops. Next, we set $\theta$ to 0.1, and vary the $QI$ size. As the $QI$ size increases, the data tend to be more sparse in $QI$-space, hence it is more likely that ECs with bigger bounding boxes are created. Thus, in Figure 8(c) the workload error increases with $QI$ size, for *all* compared methods, while BUREL presents the most modest increase. Last, Figure 8(d) presents the results as a function of selectivity $\theta$. As $\theta$ grows, the length of the range $R_A$ for each attribute in a predicate increases. This makes the minimum bounding box of an EC more likely to be entirely contained in the query region, so the estimate becomes more accurate and the error smaller. BUREL achieves consistently better utility.

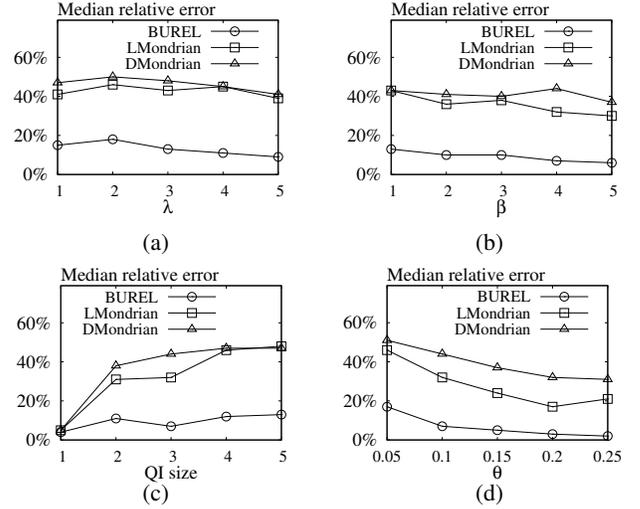

**Figure 8: Median relative error**

## 6.3 Evaluation on Perturbation

In this section we evaluate the performance of our perturbation-based $\beta$-likeness scheme discussed in Section 5. We keep the $QI$ value of each tuple unchanged, and only randomize its $\mathcal{SA}$ value according to a certain probability defined in conditional Equation 12. We emphasize that there *does not exist* information loss by generalized $QI$ values to examine as with BUREL, but we can study the utility of perturbed data set, again by aggregation queries. However, unlike the query answer estimation for generalized data using an intersection between the query and the EC, now we estimate the result simply by reconstructing the original $\mathcal{SA}$ distribution from the perturbed $\mathcal{SA}$ values of those tuples that satisfy a query's $QI$ predicates as discussed in Section 5.

Since our $\beta$-likeness scheme by perturbation is built on $(\rho_{1i}, \rho_{2i})$-privacy, for the sake of convenience we represent it as $(\rho_{1i}, \rho_{2i})$-privacy. We emphasize that, on the one hand, BUREL is based on generalization, with the desirable property of identity anonymity; on the other hand, $(\rho_{1i}, \rho_{2i})$-privacy randomizes the $\mathcal{SA}$ value of each tuple independently, and is thus immune to corruption attacks, in which one may infer the $\mathcal{SA}$ value of a victim on condition that they already know the $\mathcal{SA}$ values of some individuals [30]. However, BUREL and $(\rho_{1i}, \rho_{2i})$-privacy are mutually incomparable. Besides, there is no previous work that achieves a privacy guarantee comparable to $\beta$-likeness by perturbation; the most recent related work that offers a privacy guarantee by perturbation, [5], is also built on $(\rho_{1i}, \rho_{2i})$-privacy, yet only limits the posterior probability of inferring any individual $\mathcal{SA}$ value, a privacy guarantee comparable to $\ell$-diversity. In the absence of another competitor, we introduce and compare to a Baseline approach, which publishes the exact $QI$ value of each tuple together with the overall $\mathcal{SA}$ distribution in the original table, in the way of Anatomy [33].

Figure 9 shows our results. We first set $QI$ size to be 5, query selectivity $\theta = 0.1$, and vary the number of $QI$-attributes in the aggregation queries. The $QI$ value of each tuple remains intact for both $(\rho_{1i}, \rho_{2i})$-privacy and Baseline. Thus, only the predicate on $\mathcal{SA}$, pred($\mathcal{SA}$), incurs an error. As $\lambda$ grows, the query range interval $R_{\mathcal{SA}}$ for $\mathcal{SA}$ also increases, in effect more tuples satisfy the query, and the reconstructed $\mathcal{SA}$ distribution is closer to the actual one.

Therefore, as Figure 9(a) shows, the workload error decreases as a function of $\lambda$. Next, we set $\lambda = 3$, $\theta = 0.1$, and study the effect of $\beta$. Baseline is independent of $\beta$; the small fluctuation of its curve is due to the fact that we randomly generate $R_A$ (i.e., the query range interval for an attribute) in each experiment. However, $f(p_i)$, the allowed posterior confidence of an attacker on $\mathcal{SA}$ value $v_i$, grows as a function of $\beta$. A higher value of $f(p_i)$ implies a larger $\alpha_i$, allowing for a higher probability that an $\mathcal{SA}$ value remains intact after randomization. Therefore, the data utility rises as $\beta$ grows (Figure 9(b)). Next, we set $\beta = 4$, and vary QI size; Figure 9(c) shows the results. As neither $(\rho_{1i}, \rho_{2i})$-privacy nor Baseline modifies any QI value, the utility of perturbed data depends on the input data set. Therefore, the workload error does not change uniformly with QI size. Last, we study the effect of varying $\theta$. When $\theta$ is larger, $R_{\mathcal{SA}}$ also grows. Hence, more tuples satisfy the query, and the result becomes more accurate, as Figure 9(d) shows. Remarkably, in all presented cases, the accuracy of our perturbation-based scheme consistently outperforms that of the Baseline approach.

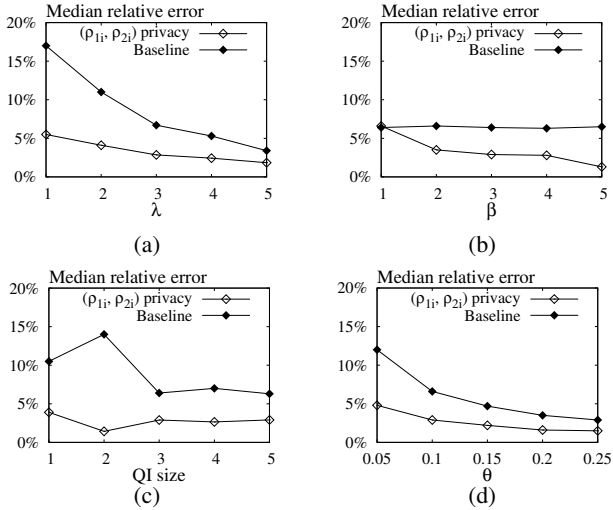

**Figure 9: Median relative error with perturbation**

## 7. RESISTANCE TO ATTACKS

We now discuss the resistance of our model and schemes to several types of attack proposed in the literature.

A *minimality attack* [31] applies when an algorithm populates ECs with tuples explicitly heeding to privacy considerations, making decisions "uniquely decided by the sensitive value of a particular tuple" [34]. BUREL decides first on EC dimensions, considering $\mathcal{SA}$ values alone. Then, it decides on the particular contents of each EC, *independently* of others, looking only at tuples' QI values and heeding to utility considerations; it does not decide whether to put a given tuple in one EC or another by looking at its $\mathcal{SA}$ value. This separation of tasks renders BUREL immune to minimality attacks. Furthermore, [7] has shown that the minimality attack can be easily averted even in the case of algorithms vulnerable to it.

A *deFinetti attack* [15] aims to learn the correlation between $\mathcal{SA}$ values and QI values by building a Bayesian network; it starts by assuming a random permutation to assign each $\mathcal{SA}$ value to a QI value in each EC, and builds a Naïve Bayes classifier out of all such assignments. Then it evaluates the permutation assigned to each EC, and generates an improved one, which is in turn used to update the classifier. This iterative procedure goes on until it converges. In other words, the classifier exploits divergences between the *global information*, as it appears in the whole published table, and *local knowledge* within each EC, to iteratively correct the $\mathcal{SA}$ to QI assignments within each EC. We deduce that, if this divergence is controlled, the success potential of the deFinetti attack can be correspondingly constrained. The $\beta$-likeness principle delimits exactly this divergence by a threshold $\beta$, hence constrains how much an attacker learns beyond the overall distribution in a published table. We thus argue that $\beta$-likeness curbs the deFinetti attack as the value of $\beta$ prescribes. Intuitively, a lower $\beta$ value allows for smaller divergences and hence lower success rate of the attack. We have defined $\beta$-likeness in a way that constrains positive, but not negative, information gain, as this is the cardinal need in most practical circumstances. Still, a deFinetti attack may also exploit negative divergences in order to construct its classifier. In case such concerns arise, our model can be straightforwardly extended to constrain negative divergences as well, and thereby further enhance its capacity to thwart such attacks.

Cormode [6] recently conducted an experimental study of the deFinetti attack on Anatomy [33], an instantiation of $\ell$-diversity, concluding that the attack is effective for small values of $\ell$ (2, 3, 4). Still, as $\ell$ rises, the attack's success rate deteriorates. In particular, for $\ell = 5$ the rate is below 50%, and when $\ell$ reaches 7 it falls below 30%. As the attack has so far only been implemented against Anatomy, presenting the privacy of data anonymized by BUREL in terms of $\ell$-diversity is relevant in this context. The table on the right presents the $t$ and $\ell$ values achieved in terms of $t$-closeness and $\ell$-diversity, respectively, for the data sets published by $\beta$-likeness in the exper-

| $\beta$ | $t$ | Avg $t$ | $\ell$ | Avg $\ell$ |
|---|---|---|---|---|
| 1 | 0.02 | 0.01 | 19.0 | 20.7 |
| 2 | 0.09 | 0.04 | 11.0 | 15.9 |
| 3 | 0.13 | 0.04 | 8.7 | 14.2 |
| 4 | 0.16 | 0.04 | 7.2 | 13.6 |
| 5 | 0.17 | 0.05 | 6.6 | 12.6 |

iment of Figure 4(a), with $\beta$ set to 1, 2, 3, 4, and 5; Avg $\ell$ ($t$) stands for the average diversity (closeness) for all the ECs. Notably, for reasonable values of $\beta$, $\ell$ assumes values no less than 6 for which the deFinetti attack's succcess rate is low.

The hitherto discussed attacks are designed against generalization-based schemes. Our perturbation scheme is not vulnerable to them, as it involves no generalization. Moreover, as it randomizes each $\mathcal{SA}$ value independently, it is immune to *corruption attacks* [30], in which an attacker who is already aware of the $\mathcal{SA}$ values of some individuals tries to infer that of a victim. Besides, our schemes assume the anonymized data are published only once, so as to prevent *composition attacks* [11]. Thwarting such attacks with republication under $\beta$-likeness is a problem orthogonal to our work.

Cormode [6] also suggests an attack on differential privacy based on a Naïve Bayes classifier. Such a classifier predicts the $\mathcal{SA}$ value of a tuple $t$ with $m$ QI-attribute values, $t_j$, $1 \leq j \leq m$, as:

$$\hat{v}(t) = \underset{v_i \in \mathcal{V}}{\operatorname{argmax}} \Pr[v_i] \prod_{j=1}^{m} \Pr[t_j|v_i] \quad (15)$$

The gist of the attack lies in the fact that the conditional probabilities $\Pr[t_j|v_i]$ can be accurately learned based on noisy count query results extracted from differentially private data. While the noise in question conceals the contribution of any individual, its effect on the derived $\Pr[t_j|v_i]$ is relatively small [6]; thus, the built classifier works almost as effectively as in the noiseless case, exploiting variations of $\Pr[t_j|v_i]$ values from their unconditional counterpart, $\Pr[t_j]$ to produce a non-trivial prediction of $\hat{v}(t)$. On the other hand, $\beta$-likeness is defined in a way that explicitly bounds exactly the variation of these conditional probabilities from their unconditional counterpart. Specifically, by Bayes' rule, we get:

$$\Pr[t_j|v_i] = \frac{\Pr[v_i|t_j]}{\Pr[v_i]} \Pr[t_j] \quad (16)$$

For a given sensitive value $v_i \in \mathcal{V}$, $\Pr[v_i]$ is the prior confidence in $v_i$ based on the global distribution of $\mathcal{SA}$ values, which we have hitherto denoted as $p_i$, while $\Pr[v_i|t_j]$ is the posterior confidence that $\beta$-likeness bounds by $f(p_i) = (1 + \min\{\beta, -\ln p_i\}) \cdot p_i$. Then $\beta$-likeness guarantees that $\Pr[t_j|v_i] \leq (1 + \min\{\beta, -\ln p_i\}) \cdot \Pr[t_j]$.
1398

Thus, $\beta$-likeness bounds the conditional probabilities that the Naïve Bayes attack exploits, delimiting the extent to which their values vary from $\Pr[t_j]$. Consequently, $\beta$-likeness delimits the potential for a Naïve Bayes attack to succeed, causing Equation (15) to predict the most frequent $\mathcal{SA}$ value in the table most of the time.

The preceding analysis has been made without prejudice to the publication format, and hence applies to any scheme satisfying $\beta$-likeness. However, the same analysis can be made specifically for publication by generalization. Assume BUREL outputs $e$ ECs, and $f$ of those include $QI$ attribute value $t_j$. Let $\{G_1, \ldots, G_f\}$ be the set of ECs that contain $t_j$, $\{G_{f+1}, \ldots, G_e\}$ the set of all other ECs, and $q_i^k$ the frequency of $\mathcal{SA}$ value $v_i$ in EC $G_k$. Then it is:

$$\Pr[t_j|v_i] = \frac{q_i^1|G_1| + q_i^2|G_2| + \ldots + q_i^r|G_f|}{p_i \cdot (|G_1| + \ldots + |G_e|)} \quad (17)$$

$$\leq (1 + \min\{\beta, -\ln p_i\}) \cdot \frac{|G_1| + \ldots + |G_f|}{|G_1| + \ldots + |G_e|} \quad (18)$$

$$\leq (1 + \min\{\beta, -\ln p_i\}) \cdot \Pr[t_j] \quad (19)$$

The last inequality confirms our previous result. For illustration, we estimate $\Pr[t_j|v_j]$ values as in Equation (17) on the anonymized CENSUS data, using the first three attributes as $QI$, to predict the $\mathcal{SA}$ value of each tuple by Equation (15), for $\beta \in \{1, 2, 3, 4, 5\}$. We obtain the success rate shown in the figure above. As expected, this success rate remains remarkably close to the frequency of the most frequent $\mathcal{SA}$ value in the data, namely $4.8402\%$.

Last, we emphasize that $\beta$-likeness is a privacy model for *categorical data*. Its extension to *numerical data* is an interesting topic for future research. Such an extension should constrain not merely the variation in the frequencies of discrete numerical *values*, but rather of any values in close *proximity* to each other. Doing so, it would be immune to *proximity attacks* [19], as they apply on numerical data. In case proximity is defined for categorical data by a *semantic hierarchy* of categorical values, our model can be easily extended so as to treat all values beneath the same selected nodes in this hierarchy as the same, and ensure $\beta$-likeness for such groups of values instead of leaf nodes in the hierarchy. We also emphasize that out model is built under the assumption that an attacker has no other prior knowledge apart from the overall distribution of sensitive values. Rastogi et al. [26] show that, if an adversary knows *arbitrary correlations* among tuples, there exists no useful anonymization algorithm that can achieve both privacy and utility.

## 8. CONCLUSION

In this paper we revisited the microdata anonymization problem with three distinct contributions. *First*, we introduced $\beta$-likeness, a robust privacy model that provides a comprehensible and intuitively appealing privacy guarantee, expressed as a limit on the relative confidence gain on *each single* sensitive attribute value. *Second*, we devised BUREL, a novel generalization algorithm explicitly customized for this model. *Third*, we devised a perturbation technique for our model. Our experimental results confirm that algorithms developed for other privacy models cannot achieve strong guarantees in terms of $\beta$-likeness, and verify the effectiveness and efficiency of both our schemes in their task. Apart from this experimental study, we also provided arguments and results to the effect that the $\beta$-likeness privacy guarantee affords genuine protection against attacks suggested in previous research. In the future, we intend to extend our model to numerical sensitive attributes.

## Acknowledgments
We thank Daniel Kifer and Graham Cormode for lucid remarks on this topic, and the anonymous reviewers for their apt feedback.